\documentclass[a4paper]{article}

\usepackage{INTERSPEECH2021}

\usepackage{amsmath,graphicx}
\usepackage{xcolor}
\usepackage{multicol}
\usepackage{multirow}
\usepackage{subfig}

\usepackage[utf8]{inputenc}
\usepackage{amssymb,epsfig,url,mathrsfs} 
\usepackage{algorithm,algorithmic}
\usepackage[framemethod=TikZ]{mdframed}
\usepackage{xcolor}
\usepackage[colorinlistoftodos]{todonotes}
\usepackage{enumitem}

\renewcommand{\vec}[1]{\boldsymbol{\mathrm{#1}}}

\newcommand{\transp}{\ensuremath{^\mathsf{T}}}

\renewcommand{\vec}{\boldsymbol} % Vector

\title{Data Quality as Predictor of Voice Anti-Spoofing Generalization}

\name{Bhusan Chettri$^{1,2}$, Rosa Gonz\'alez Hautam\"aki$^{1,4}$, Md Sahidullah$^{3}$, Tomi Kinnunen$^1$\sthanks{$\;$All authors have equal contribution.}}
\address{
$^1$School of Computing, University of Eastern Finland, Finland \\ 
$^2$School of EECS, Queen Mary University of London, United Kingdom \\
$^3$Universit\'{e} de Lorraine, CNRS, Inria, LORIA, F-54000, Nancy, France \\
$^4$Department of Electrical and Computer Engineering, National University of Singapore, Singapore}
\email{b.chettri@qmul.ac.uk, rgonza@cs.uef.fi, md.sahidullah@inria.fr, tkinnu@cs.uef.fi}
\begin{document}
%\ninept
%
\maketitle
\begin{abstract}
Voice anti-spoofing aims at classifying a given utterance either as a bonafide human sample, or a spoofing attack (e.g. synthetic or replayed sample). %Numerous voice 
Many anti-spoofing methods have been proposed but most of them fail to generalize across domains (corpora) --- and we do not know \emph{why}. We outline a novel interpretative framework for gauging the impact of data quality upon anti-spoofing performance. Our within- and between-domain experiments pool data from seven public corpora and three anti-spoofing methods based on Gaussian mixture and convolutive neural network models. We assess the impacts of long-term spectral information, speaker population (through x-vector speaker embeddings), signal-to-noise ratio, and selected voice quality features.
\end{abstract}
\noindent\textbf{Index Terms}: anti-spoofing, %presentation attack detection %, 
data quality, interpretative models

\section{Introduction}
\label{sec:intro}

In the context of biometrics, %person authentication, 
\emph{presentation attack detection} (PAD) or \emph{anti-spoofing} aims at classifying a given signal either as a bonafide (human) sample or a \emph{spoofing attack}. Replay, text-to-speech, and voice conversion attacks degrade the performance of automatic speaker verification (ASV) systems. Driven by fraud prevention in call-centers and securing our identities in other applications, a new research community working on voice anti-spoofing has emerged during the past few years. In part, research has been enabled by increased number of corpora containing both bonafide and spoofed data, such as ASVspoof \cite{Wang2020-ASVspoof2019}. There are also other public (and proprietary) data such as BTAS 2016 \cite{Korshunov-btas2016}, SAS \cite{wu-SAS}, ReMASC \cite{Gong2019-ReMASC}, and PhoneSpoof \cite{Lavrentyeva2019-phonespoof}.

\begin{figure}[t!]
	\centering 
	\includegraphics[width=0.9\linewidth]{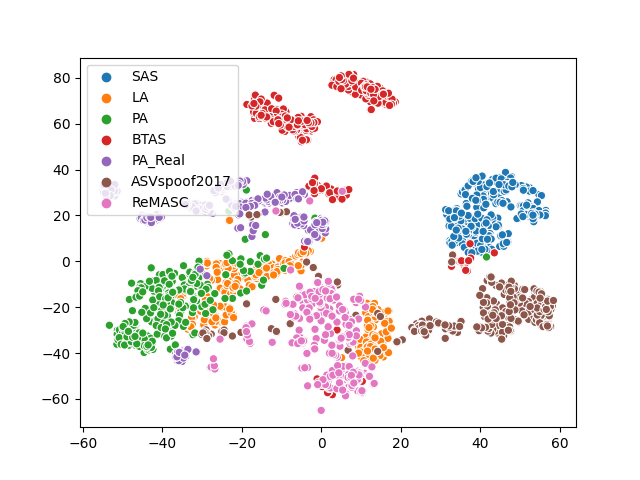}
	\vspace{-0.3cm}
	\caption{Voice anti-spoofing corpora visualized with t-stochastic neighborhood embedding (t-SNE). Each point corresponds to long-term average spectrum (LTAS) of one utterance.}
	\label{fig:tsne-visuals}
	\vspace{-0.7cm}
\end{figure}

Numerous speaker-independent voice anti-spoofing methods have been proposed. Many focus on designing new acoustic features \cite{Todisco2017-constant-Q,gajan_IS2018}, deep neural network (DNN) architectures \cite{Lavrentyeva2017-audio-replay,Gomez_RCNN_TASLP} or combining different models \cite{Chettri2019,Chen2017} through classifier fusion. Many studies report low spoofing attack detection error rates (even 0 \%) though the methods are usually tested using a single corpus only. With sufficiently many architectural modifications, control parameter optimizations and experiments it may be feasible to push %detection 
error rates down on a given corpus. Performance on a single corpus, however, should not be viewed as a measure of generality or to suggest a solved task. Real-world operation demands reliable operation across \emph{many} test conditions, most of which are never encountered during system development. 

%\cite{Tom2018}
Lack of generality has been noted in (limited number of) \emph{cross-corpus} studies \cite{Korshunov2019-cross-db,paul2017generalization,das2020assessing,parasu2020} where the training and test data originate from disjoint collections (often compiled by different research teams). The reported error rates, sometimes close to chance level, are disturbing as they suggest overfitting on the existing corpora. With a spoiler alert, the reader is encouraged to peek our cross-corpus results reported in Table \ref{tab:cross-corpus-performance-summary}. %below. %as an independent validation.

But \emph{why} voice anti-spoofing, especially across corpora, is so difficult? As an intuitive motivation, Fig. \ref{fig:tsne-visuals} visualizes spectral differences in seven different voice anti-spoofing corpora. Even if each corpus contains different spoofing attacks of varied difficulty, at the corpus level the audio files can be homogenous. This is due to shared acoustic properties that may depend on speaker population, original recording environment, choice of microphones, data processing pipelines --- and perhaps even on signal scale and audio file format. Similarly, there are systematic differences across corpora due to differences in such characteristics. When a voice anti-spoofing system is trained and tested using data in a single corpus only, one conveniently sidesteps the issue of feature or representation compatibility across domains; it may not be needed as the training and test data are already homogenous in their qualities.

Our work aims at quantifying the impact of corpus-level acoustic mismatch factors upon voice 
anti-spoofing performance. Our work is differentiated from majority of prior work in anti-spoofing by an \emph{explanatory} perspective. As a community, we lack \emph{understanding} of the role of training and test data in %voice 
anti-spoofing. Given the central role of data in any machine learning task (including anti-spoofing), we argue that it is useful to uncover data-related factors that contribute negatively (or positively) to %voice anti-spoofing 
performance. We approach this problem by focusing on a few carefully selected corpus level attributes, such as distribution of signal-to-noise ratio and speakers. These potentially confounding variables are then used as predictors of anti-spoofing performance in a regression analysis setting.

Our work is not the first to address the impact of factors that may influence anti-spoofing performance or bias evaluation results. Prior work has addressed, for instance, the impact of waveform sample distributions \cite{Lapidot2020} and biases due to presence of silence regions \cite{Chettri2019,bhusan_taslp}. Other work, such as \cite{Tak2020}, have provided interpretations beyond error rates for specific anti-spoofing methods. Our work is differentiated from these studies in that we propose a \emph{unified} framework for assessing data-related quality factors, treated as predictors in a regression model setting. %This involves random sampling of several training-test splits for each corpus. 
What follows is description of our framework and preliminary experiments that pool data from seven different anti-spoofing corpora.

\section{Methodology}\label{sec:methodology}

\subsection{Re-thinking training and test sets as random data}

Assume that we have a total of $M$ distinct, labeled anti-spoofing collections $\{\mathcal{D}_i\}_{i=1}^M$ available (here, $M=7$). The $i^\text{th}$ collection contains, respectively, $N_\text{bona}^{(i)}$ and $N_\text{spoof}^{(i)}$ bonafide (human) and spoof audio files. Each file is labeled as either one of these two classes. %If the recordings across different corpora are distinct, there are a total of $N_\text{bona}=\sum_{i=1}^M N_\text{bona}^{(i)}$ and $N_\text{spoof}=\sum_{i=1}^M N_\text{spoof}^{(i)}$ unique bonafide and spoof recordings, respectively.
Each collection (e.g. particular ASVspoof edition) is assumed to consist of somewhat homogenous audio material, while different collections --- possibly compiled by different researchers --- are assumed to be more heterogenous. Each %of the 
collection can be understood as a cluster or group of audio files that share some commonalities. The reported performance gap of within-corpus vs. cross-corpus results \cite{Korshunov2019-cross-db}, along with Table \ref{tab:cross-corpus-performance-summary} and the visualization in Fig. \ref{fig:tsne-visuals} on long-term spectral characteristics provide support for these assumptions.

Typically, a voice anti-spoofing corpus contains an \emph{evaluation protocol} that defines partitioning of the speech files into training and test portions\footnote{ASVspoof challenges contain \emph{train}, \emph{development} and \emph{evaluation} sets; we do not differentiate between the latter two which, really, are two different test sets. %The difference is that 
During a challenge, the labels of development set are available for detector optimization while test data that lacks labels.}. %needs to be processed blindly.}. 
Even if standard evaluation protocols are necessary for commensurable performance comparisons, a protocol defines only one possible data partitioning of all the available data. As a result, reported anti-spoofing results on a given corpus may be specific to that random partitioning. In stark contrast to fixed train-test protocol division, \emph{we consider the training/test corpora as random observations}. Whenever the anti-spoofing system (and its parameters) are frozen, one obtains \emph{one} performance number (such as equal error rate, EER) for a fixed evaluation protocol. We, instead, gather \emph{several} repeated measurements of the selected performance measure (here, the EER) within and across data collections.

In practice, for each of the $M$ collections we designate a \emph{single} training set $\mathcal{D}_\text{train}^{(i)}$ and \emph{multiple} test sets, $\mathcal{D}_\text{test}^{(i,j)},j=1,\dots,N_\text{test}^{(i)}$. In principle, this choice is arbitrary and we could have also fixed the test sets and sample random training sets instead. The choice is primarily dictated by computational reasons elaborated shortly. We sample equal number of test portions within each collection:  $N_\text{test}^{(1)}=\dots=N_\text{test}^{(M)} \equiv N_\text{test}$. Note that the special case $N_\text{test}=1$ corresponds to conventional approach where a given corpus is equipped with a pre-defined evaluation protocol. In our revised set-up we train and test anti-spoofing systems across all the collections. This yields $N_\text{test}$ within-corpus and $(M-1) \times N_\text{test}$ cross-corpus experiments, \emph{per training set}. As we have $M$ training sets (one per corpus), we have a total of $M \times N_\text{test}$ within-corpus results and $M \times (M-1) \times N_\text{test}$ cross-corpus results. In our experiments, $N_\text{test}=20$ which implies 140 within- and 840 cross-corpus experiments. This is %the reason 
why we fix the training partition and treat only the test portions as random: despite the large number of EERs produced, we need to train only $M=7$ anti-spoofing models (one per collection).

\begin{table}
	\caption{Cross-corpus performance (EER\%) of spoofing countermeasures. 2017: ASVspoof 2017 v2.0, PA: ASVspoof 2019 PA, RPA: ASVspoof 2019 Real PA, LA: ASVspoof 2019 LA. ReM: ReMASC, BT: BTAS. An EER of greater than $50$\% indicates chance level in a 2-class task.} %that spoofed speech score distribution has shifted towards positive side compared to that of human speech.}}
	
	\centering
	\vspace{-0.2cm}
	\scalebox{0.8}{
	\begin{tabular}{c|ccccccc}
	\hline
	& \multicolumn{7}{c}{Tested on} \\
	\cline{2-8}
	&SAS &LA &2017 &RPA &ReM &PA &BT\\
	\hline
	SAS &$\textbf{0.99}$ &$62.08$ &$53.76$ &$70.18$ &$46.85$ &$52.29$ &$73.51$\\
	LA  &$45.07$ &$\textbf{11.17}$ &$41.06$ &$34.0$ &$49.37$ &$36.16$ &$81.75$\\
	%\hline
	2017 &$52.97$ &$39.07$ &$\textbf{13.02}$ &$41.83$ &$43.52$ &$47.92$ &$70.6$\\
	RPA &$52.01$ &$53.75$ &$46.16$ &$\textbf{39.35}$ &$45.46$ &$46.52$ &$54.77$\\
	ReM &$42.27$ &$24.2$ &$54.08$ &$58.58$ &$\textbf{50.3}$ &$48.88$ &$66.02$\\
	                              
	%\hline
	PA &$65.56$ &$24.94$ &$52.21$ &$29.88$ &$47.39$ &$\textbf{7.0}$ &$13.87$\\
	%\hline
	BT &$61.13$ &$68.59$ &$17.03$ &$33.49$ &$43.78$ &$46.44$ &$\textbf{0.18}$\\
	\hline
	\end{tabular}}

	\label{tab:cross-corpus-performance-summary}
	\vspace{-0.4cm}
\end{table}

\subsection{Overview of multiple linear regression setting}

%Our work 
We model the dependency of anti-spoofing performance upon data-related mismatch factors. For instance, if the training and test data consist of homogenous speakers (e.g. all have the same gender or native language) one might expect better performance compared to a situation with disjoint speaker qualities. % Similarly, if both the training and test data are of similar quality (e.g. both are clean; or both are noisy) one may obtain higher accuracy.
We consider paired observations $\{(\vec{d}_t, E_t): t=1,\dots,T\}$ where $E_t$ is performance metric (here, bonafide-vs-spoof EER) for training-test pair indexed by $t$ and $\vec{d}_t = (d_t^{(1)},\dots,d_t^{(R)})\transp \in \mathbb{R}^R$ is a set of predictors suspected to influence $E_t$. We model the assumed statistical dependency using \emph{multiple linear regression}.
%a regression model $E_t = f(\vec{d}_t|\vec{\theta}) + \varepsilon_t$ where $f(\vec{d}_t|\vec{\theta})$ is a deterministic function  and $\varepsilon_t \sim_\text{i.i.d.} p(\varepsilon|\vec{\lambda})$ is a random residual. Any regressor could be used but we use linear models for the sake of interpretability. %This is the setup of standard \emph{multiple linear regression} analysis. 
Our prime interest is in the relative contribution of the individual predictors $d_t^{(1)},\dots,d_t^{(R)}$, each of which is a distance %or divergence 
between the training and test sets, formalized next.

\subsection{Defining the predictors (corpus distances)}

Let $\mathcal{D}_\text{train}$ and $\mathcal{D}_\text{test}$ denote training and test sets that are used, respectively, to train and score any anti-spoofing system. They could be sets within the same collection or sets taken from different collections; this distinction is not important as the procedure of distance computation is the same. Let $\mathcal{D}_\text{train}=\{(\mathcal{X}_j, y_j)\}_{j=1}^{N_\text{train}}$ and $\mathcal{D}_\text{test}=\{(\mathcal{X}_m, y_m)\}_{m=1}^{N_\text{test}}$ denote training and test waveforms $\mathcal{X}$ paired up with their ground-truth labels, $y \in \{0\equiv\text{spoof}, 1\equiv\text{bonafide}\}$. The $j$th waveform, $\mathcal{X}_j$, is represented by a set of quality features, $\vec{\phi}_j^{(1)},\dots,\vec{\phi}_j^{(Q)}$. %Each of the $Q$ features 
They may have different dimensionalities and numerical ranges. For instance, $\vec{\phi}_j^{(1)}$ might be scalar-valued signal-to-noise ratio (SNR) and $\vec{\phi}_j^{(2)}$ a 512-dimensional deep speaker embedding. Each feature set corresponds to attributes suspected to influence  anti-spoofing performance but which (ideally) should be uninformative about the class label $y$. For instance, one is not supposed to detect a spoofing attack based on knowledge of the speaker (at least in speaker-independent anti-spoofing setting). At the level of the corpus, however, it is useful to gauge the potential impact of speaker population upon anti-spoofing performance. 

\begin{figure}
    \centering
    \includegraphics[scale=0.35]{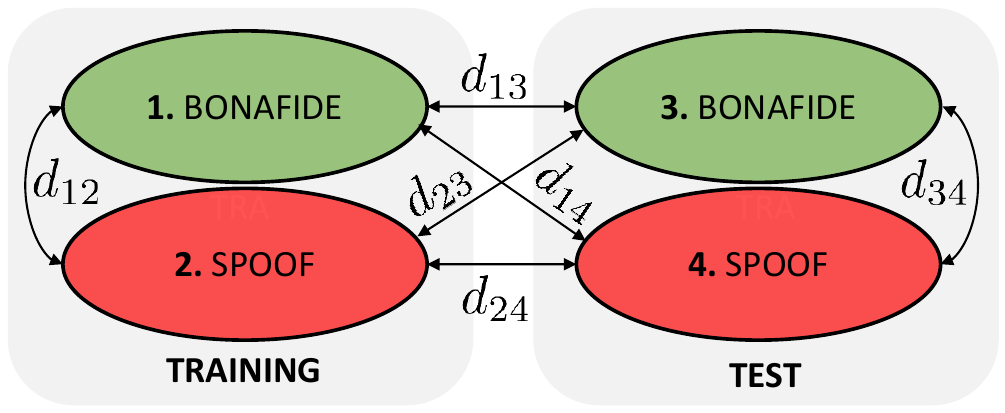}
    \caption{For each of $Q$ quality measures, six distances are computed: within- and between-class distances of bonafide and spoof (both within and across training and test data).}
    \label{fig:distance-illustration}
    \vspace{-0.5cm}
\end{figure}

In practice, we treat each of the $Q$ features independent of each other. %design custom distance measure for each case. There are, however, some general principles to be followed. Since each quality measurement is treated independent of each other, 
We drop the feature superscript momentarily and use $\vec{\phi}_j$ to denote any of the $Q$  measurements of file $j$. The observed quality data are then
$\mathcal{D}_\text{train}=\{(\vec{\phi}_j, y_j)\}_{j=1}^{N_\text{train}}$ and $\mathcal{D}_\text{test}=\{(\vec{\phi}_m, y_m)\}_{m=1}^{N_\text{test}}$, viewed as i.i.d. samples from some underlying true distribution $p(\vec{\phi},y)$. %Alternatively, by conditioning on the class label, $\vec{\phi}_j | \text{bonafide} \sim_\text{i.i.d.} p(\vec{\phi}|y=$
%We introduce another categorical variable $z \in \{\texttt{train}=0,\texttt{test}=1\}$ to indicate the data portion. 
By conditioning the data distribution both by the class label (bonafide/spoof) and the data portion (train/test) we have four conditional data distributions in total, as illustrated in Fig. \ref{fig:distance-illustration}. %For instance, $p(\vec{\phi}|y=0,z=0)$ denotes the training set distribution of spoof data. 
%If the class label $y$ is treated as a latent variable and integrated out, both the the training and test distributions are mixture distributions (with two components and mixing weights determined by the proportion of bonafide and spoof classes). We could therefore define a single training-test distance as some distance measure between the two mixture distributions. We take a different route, however, by 
For regression modeling, we consider all the indicated six distances, for a given set of quality measurements (thus, the maximum number of predictors is $R=6Q$, obtained by cross-combining all $Q$ quality measurements with the six different distances). The distance that we use is \emph{Chamfer distance} (or \emph{modified Hausdorff distance}) based on averaged Euclidean squared distances with nearest-neighbor rule. It can be computed without numerical issues for features of any dimensionality. It gives non-negative distance of two point clouds, each of which corresponds to one the four portions shown in Fig. \ref{fig:distance-illustration}. Chamfer distance is not symmetric but we compute distance to both directions and average the two values. We also normalize the distance by the dimensionality of the respective quality measurement.

The within-class distances across training and test are perhaps most easily intuitively understood. For instance, $d_{13}$ measures how much bonafide data qualities between training and test data differ (likewise for spoof, $d_{24}$). The remaining four \emph{cross-class} distances may appear strange at first but we have a reason to include them in our models. If \emph{both} bonafide and spoof are corrupted by similar nuisance variations (e.g. both are either clean or noisy) one may expect lower anti-spoofing EER as the classifier does not have to address the issue of noise. Similarly, if the training distributions of bonafide and spoof distributions are very different, the anti-spoofing system may learn to \emph{cheat} (take a shortcut) by extracting information unrelated to bonafide-spoof discriminating cues --- hence, potentially exhibit low generalization performance.

\section{Experimental Setup}

\subsection{Spoofing corpora}
We use seven publicly available corpora: SAS \cite{wu-SAS}, ASVspoof 2017 v2.0 \cite{hectorAsvspoof2.0}, ASVspoof 2019 (LA, PA and PA Real) \cite{asvspoof2019overview}, BTAS 2016 \cite{Korshunov-btas2016}, and ReMASC \cite{Gong2019-ReMASC}. The SAS corpus was created for anti-spoofing research with %synthetic speech created with 
seven voice conversion (VC) and three speech synthesis (SS) methods. The subsequent ASVspoof 2015 corpus 
%was designed with 
includes the same attacks. %used in SAS corpus. 
While the ASVspoof 2017 corpus contains real replay attack recordings, %collected under real conditions, 
the ASVspoof 2019 PA corpus %was created in a %controlled setting 
consists of simulated replay attacks. PA real is a small test set that contains real replayed audio files. %collected in realistic replay attack conditions. 
ReMASC \cite{Gong2019-ReMASC} is a another publicly available corpus for replay spoofing attack research in voice controlled applications. We also include %use 
the% speech 
corpus used in BTAS 2016 anti-spoofing competition. It consists of different types of replay attacks~\cite{Korshunov-btas2016}. 

\subsection{Random training-test protocol design}
%To study predictors' effect modelling, 
We have created multiple train-test conditions with smaller subsets. We sampled the training data to create a smaller training subset balanced according to the number of utterances and speakers. We %have considered 
include five speakers from each corpus, each with 10 bonafide and 50 spoofed utterances. This results to 300 training utterances per corpus. Similarly, we created 20 test sets for each of the seven corpora, each consisting of 50 bonafide and 250 spoofed utterances. The bonafide-to-spoof utterance ratio %between the number of bonafide and spoofed utterance 
approximately corresponds to the ratio in standard evaluation protocols --- there are typically far more spoofed than bonafide utterances available. %For the random train and test subsets, 
We selected the speakers and the utterances from the respective pre-defined `train’ and `evaluation’ partition randomly. %respectively. 
Due to unavailability of the speaker partitioning of training and evaluation in ReMASC and ASVspoof 2019 Real PA, we select the speakers of train and test in a disjoint manner.

\subsection{Classifiers and performance measures}
We use Gaussian mixture model (GMM) and convolutional neural network (CNN) as classifiers, due to their extensive use in anti-spoofing research~\cite{Lavrentyeva2017-audio-replay,gajan_IS2018,Wang2020-ASVspoof2019,sahidullah2019introduction}. The GMM-based systems are the same as the two baseline systems used in the ASVspoof 2019 challenge. They operate on 60-dimensional linear frequency cepstral coefficients (LFCCs) and 90-dimensional constant-Q cepstral coefficients (CQCCs), respectively. Two GMMs are trained to model the distribution of bonafide and spoof data using $512$ mixture components. The CNN system, in turn, uses power spectrogram inputs. It is trained discriminatively to optimise cross-entropy between bonafide and spoofed class using Adam optimiser. We use the CNN architecture, training and testing approach from \cite{bhusan-odyssey2020}.

We evaluate classifier performance using equal error rate (EER) as a measure of bonafide-spoof discrimination. We compute EER using the public scoring toolkit used in the ASVspoof 2019 challenge. Table \ref{tab:cross-corpus-performance-summary} summarises the cross-corpus performance %\footnote{For simplicity, we only used CNN to get a high-level insight on cross-corpus performance evaluation.} 
evaluation of CNN countermeasure. %As can be seen, the models show good performance on the same corpus but perform poorly in cross-corpus scenario, as expected from prior studies \cite{Korshunov2019-cross-db}. 
As can be seen, performance is reasonable for (some) within-corpus tasks but consistently low in cross-corpus scenarios, as expected \cite{Korshunov2019-cross-db}.

\subsection{Quality features}

We include five types of quality features. Four of them are computed with rule-based methods available in common toolkits, while one (x-vector) uses a data-driven approach, which makes feature values dependent on the training data of the extractor.

\vspace{1ex}
\noindent \textbf{LTAS} represents spectral information averaged over time. We compute $257$-dimensional LTAS %vector 
per utterance using $512$-point FFT from $32$ ms Hanning-windowed frames shifted by $10$ ms.

\noindent \textbf{SNR} is computed using \emph{waveform amplitude distribution analysis} (WADA) method~\cite{kim2008robust}, which assumes that the amplitude of the speech can be approximated with Gamma distribution with shape parameter 0.4 and the noise by Gaussian distribution. WADA shows competitive performance compared to the DNN-based data-driven methods specially in higher SNR conditions~\cite{li2020frame}, case relevant for our data.

\noindent \textbf{Noise spectrum} Besides scalar-valued SNR, we also estimate noise spectral density using optimal smoothing and minimum statistics method~\cite{martin2001noise}. The method estimates noise for all frequency bins in every speech frame. We average these noise spectral densities %by taking average over all the frames resulting 
to obtain 257 coefficients per utterances.  

\noindent \textbf{X-vector} represents 512-dimensional deep speaker embedding extracted with pre-trained models trained on VoxCeleb corpus~\cite{kaldirecipe}, processed further with length normalization. %before computing the corpus distances. 
Though x-vectors depend on training data and may contain nuisance variations~\cite{raj2019probing}, the pre-trained model shows reasonable speaker verification EER of $3.13\%$ on VoxCeleb1 test set. This indicates high specificity to speaker-related information.  

\noindent \textbf{Acoustic descriptors} are extracted using openSMILE toolkit 2.3~\cite{openSMILE2013}: \emph{fundamental frequency} (F0), \emph{formant frequencies} (F1 to F4), and \emph{loudness}. The feature extraction configuration corresponds to the \emph{extended Geneva
Minimalistic Standard Parameter Set}~\cite{egemaps_2016} summarized with the \textbf{mean} of the descriptor at the utterance level. F0 is presented in a %logarithmic scale (semitones). %in a semitone frequency scale at 27.5 Hz. 
semitone scale. %Formants are the center frequency for first to fourth formants. 
Loudness is an estimate of the perceived signal energy from an auditory spectrum %that follows a 
from perceptual linear prediction (PLP) analysis~\cite{hermansky1990perceptual}.

\section{Results}
\begin{figure}[t!]
	\centering    
	\includegraphics[width=0.80\linewidth]{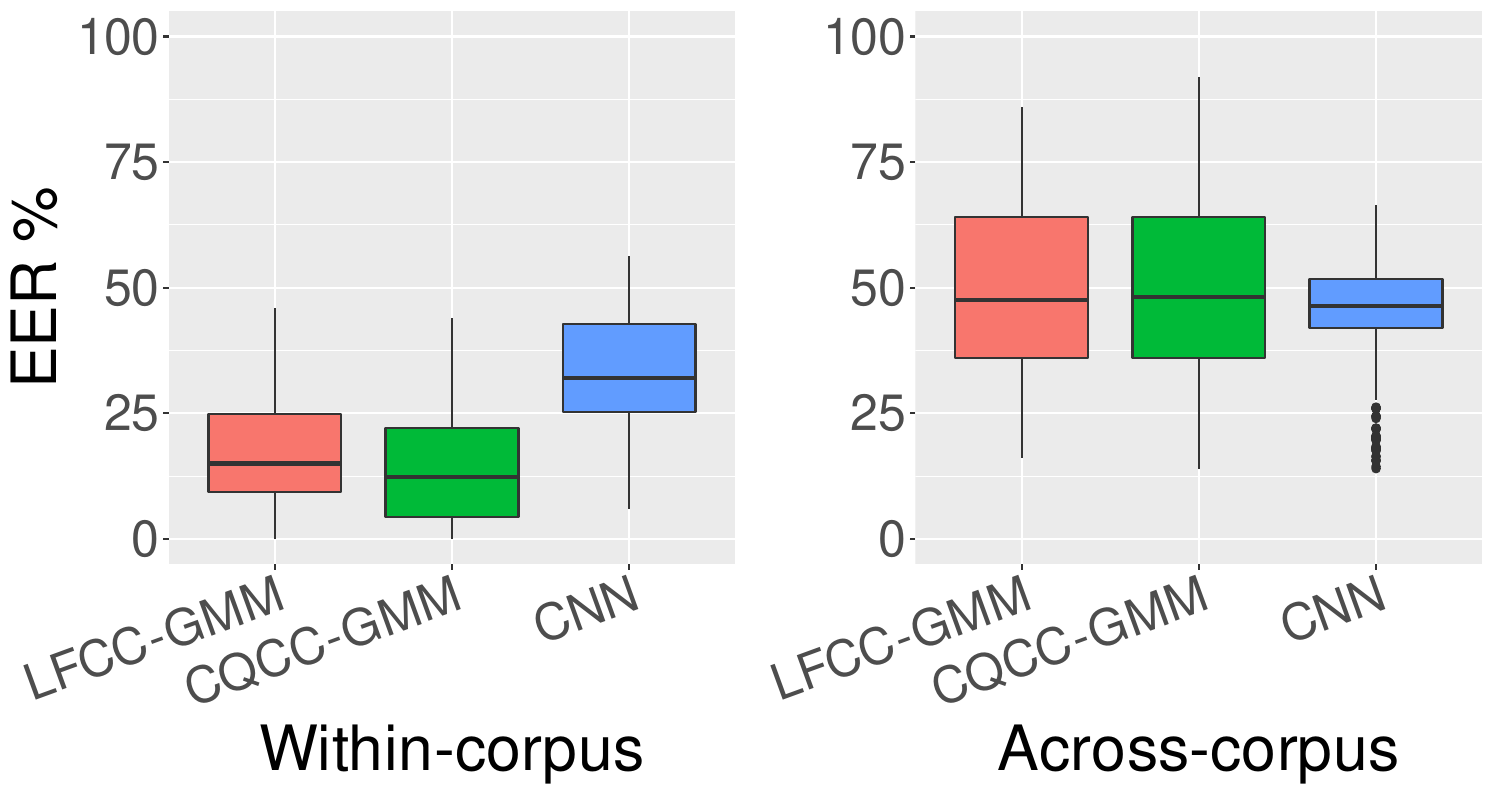}
	 \vspace{-0.2cm}
	\caption{EER distribution on 20 randomly created trial lists.}
	\label{fig:eers}
	 \vspace{-0.5cm}
\end{figure}

We explore the collinearities between the predictive features and the performance of classifiers using the Pearson correlation. The relation was analyzed considering the data grouped in within- and across-corpus that include 140 and 840 data points respectively, each with the six distances from the predictive features (as illustrated in Fig.~\ref{fig:distance-illustration}). Figure~\ref{fig:eers} shows the EER distribution for the three classifiers and describes the dependent variable variations to be explored by the regression models.

Table~\ref{tab:ltas_distances_cor} shows, as an example, the correlation of %individual 
LTAS feature distances with the EER of the CNN classifier (similar trends were observed for the other predictive features). 
Within-corpus correlations are stronger than across-corpus correlations. This indicates collinearity of the within-corpus distances and the performance of the classifiers. As for the distances, the four cross-class distances ($d_{12}, d_{14}, d_{23}, d_{34}$) have stronger correlations with EER (whether positive or negative) than the within-class distances ($d_{13}, d_{24}$). Note also that, apart from $d_{13}$ on across-corpus case, the within-class correlations are positive. This is as expected: the larger the domain mismatch in either bonafide or spoof class, the higher the EER.

\begin{table}[thb!]
\caption{Pearson correlation between LTAS distances and the equal error rate for the CNN classifier.}
\vspace{-0.2cm}
\scalebox{0.75}{
\begin{tabular}{lc|c|c|c|c|c|}
\cline{2-7}
%& \multicolumn{6}{|c|}{\textbf{Adjusted R-squared for CNN EER}}\\ \cline{2-7}
& \multicolumn{1}{|c|}{$d_{12}$} & \multicolumn{1}{c|}{$d_{13}$} & \multicolumn{1}{c|}{$d_{23}$} & \multicolumn{1}{c|}{$d_{14}$} & \multicolumn{1}{c|}{$d_{24}$} & \multicolumn{1}{c|}{$d_{34}$}\\ \hline
\multicolumn{1}{|c|}{Within-corpus} & $-0.605$ & $0.162$ & $-0.584$ & $-0.651$ & $0.367$& $-0.737$ \\ \hline
\multicolumn{1}{|c|}{Across-corpus} & $0.107$ & $-0.115$ & $0.044$ & $-0.107$ & $0.029$ & $0.099$ \\ \hline
\end{tabular}}
\vspace{-0.1cm}
\label{tab:ltas_distances_cor}
\end{table}

We now turn our focus on the predictive features. To this end, we created \emph{multiple linear regression model} for each feature to measure how well the six distances predict the corresponding EER. %combine all the six distances for predicting the classifier's EER. The aim is to measure how well the feature distances predict the corresponding EER. 
The \emph{coefficient of determination}, or $R^2$~\cite{casella2002statistical}, measures the proportion of the total variation of the dependent variable (EER) that is explained by the fitted model. The higher the number, the better the model fits the data. \emph{Adjusted}-$R^2$ takes into account the number of predictors included in the model and how they contribute information. If the predictor is not significant, the adjusted-$R^2$ will compensate it by penalizing the model fit.

\begin{table}[thbp!]
\caption{Adjusted-$R^2$ for grouped feature distances models of within- and across-corpus data of the three classifiers. Dimensionality of each feature set is indicated in parenthesis.}
\vspace{-0.2cm}
\scalebox{0.8}{
\begin{tabular}{|l|ccc|ccc|}
\cline{2-7}
 \multicolumn{1}{c}{}& \multicolumn{3}{|c|}{\textbf{Within-corpus data}} & \multicolumn{3}{|c|}{\textbf{Across-corpus data}}  \\ \cline{2-7}
 \multicolumn{1}{c|}{} & \textbf{LFCC} & \textbf{CQCC} &\textbf{CNN} & \textbf{LFCC} & \textbf{CQCC} &\textbf{CNN} \\
 \multicolumn{1}{c|}{} & \textbf{GMM} & \textbf{GMM} &\textbf{} & \textbf{GMM} & \textbf{GMM} &\textbf{} \\ \hline
LTAS (257) & $0.679$ & $0.543$ & $0.670$ & $0.289$ & $0.166$ & $0.038$ \\ 
F1..F4 (4) & $0.641$ & $0.497$ & $0.470$ & $0.131$ & $0.075$ & $0.142$ \\ 
F0 (1) & $0.513$ & $0.328$ & $0.073$ & $0.058$ & $0.082$ & $0.096$\\ \hline
x-vec. (512) & $0.558$ & $0.642$ & $0.817$ & $0.067$ & $0.149$ & $0.202$ \\ \hline
SNR (1) & $0.593$ & $0.715$ & $0.187$ & $0.160$ & $0.227$ & $0.075$ \\ 
Noise s. (257) & $0.649$ & $0.439$ & $0.565$ & $0.141$ & $0.207$ & $0.010$ \\ \hline
Loudness (1) & $0.455$ & $0.304$ & $0.448$ & $0.021$ & $0.050$ & $0.060$ \\ \hline
\end{tabular}}
\label{table:models_adjrsquared}
\vspace{-0.3cm}
\end{table}

Table~\ref{table:models_adjrsquared} presents the adjusted-$R^2$ for the feature models for each classifier separately for within- and across-corpus data. The values 
%The adjusted-$R^2$ 
can be compared across the rows for each classifier to identify the data quality feature that better explain the EER variations. For instance, in the within-corpus data, for LFCC-GMM classifier all the feature distance models are good at explaining the performance, particularly LTAS distance predictors explain 68\% of the EER's variation. Similar strong dependencies are noted with SNR for CQCC-GMM and with x-vector for CNN. Though the adjusted-$R^2$ are lower for across-corpus data, the same features explained the classifiers' EERs with high levels of significance. It is worth noting that our aim is to identify features that best explain the classifiers' performance, rather than searching for the best combination of different predictors. All our features explain well the variation in EER, especially for within-corpus data.  

So, what does Table \ref{table:models_adjrsquared} suggest? Due to space reasons, we arbitrarily pick the strongest and weakest individual predictors per classifier:

    \begin{enumerate}
        \item LFCC-GMM is most strongly impacted by LTAS, least by loudness;
        \item CQCC-GMM is most strongly impacted by SNR, least by loudness;
        \item CNN is most strongly impacted by x-vector, least by F0 (within-corpus) or noise spectrum (across-corpus).
    \end{enumerate}

So one may conjecture, for instance, that the CQCC-GMM system is potentially sensitive to noise  (suggested earlier through simulated additive noise experiments \cite{Hanilci2016-noisy-spoof}) and the CNN system potentially more strongly impacted by the choice of speaker population. The authors emphasize \emph{potentially}: it is acknowledged that, despite speaker-discriminative training objective, x-vectors are not `pure' speaker representations \cite{raj2019probing}. Their quality depends on several factors (including the choice of training data). 

\section{Conclusions}
\label{sec:conclusion}

We addressed the role of data quality in voice anti-spoofing generalization. The framework can be used to address statistical dependency between selected quality features and anti-spoofing performance. Pinpointing the potential issues can be used to design better anti-spoofing systems where the unwanted variations are suppressed in explicit ways.
Our future plans include addressing further quality features and distance measures, \emph{mixed effects} regression modeling, adding more powerful classifiers, and using the acquired knowledge to improve selected classifiers. In many classification tasks, performance can be improved by using additional training or adaptation data from the target domain. Our assumption, however, is that only a single training domain is available. The intention was to address `the truly unknown' in terms of domain variation. Our findings indicate that substantial further research remains in this area.
\section{Acknowledgements}
\vspace{-0.1cm}
This work was supported in part by the Academy of Finland (Proj. No. 309629) and Nokia Foundation. 
\vspace{-0.3cm}
%\vfill\pagebreak

\bibliographystyle{IEEEtran}
\bibliography{references}

\end{document}